\documentclass[prl,showpacs,twocolumn,floatfix,footinbib]{revtex4}

\usepackage{amsmath,amssymb}
\usepackage{graphicx}
\usepackage[colorlinks]{hyperref}

\begin{document}

\title{Stable propagation of a modulated particle beam in a crystal channel}

\author{A.~Kostyuk}
\altaffiliation[Also at: ]
{Bogolyubov Institute for Theoretical Physics, Kyiv, Ukraine}
\email[E-mail address: ]{kostyuk@fias.uni-frankfurt.de}
\author{A.V.~Korol}
\altaffiliation[Also at: ]
{Department of Physics,
St Petersburg State Maritime Technical University,
St Petersburg, Russia}
\email[E-mail address: ]{a.korol@fias.uni-frankfurt.de}
\author{A.V.~Solov'yov}
\email[E-mail address: ]{solovyov@fias.uni-frankfurt.de}
\author{Walter Greiner}
\affiliation{
 Frankfurt Institute for Advanced Studies,
 Johann Wolfgang Goethe-Universit\"at, 
Ruth-Moufang-Str.~1, 60438 Frankfurt am Main, Germany}

\begin{abstract}
The propagation of a modulated beam of charged particles in a planar crystal channel
is investigated. 
It is demonstrated that the beam preserves its modulation at sufficiently large
penetration depths  to ensure the feasibility
of using a crystalline undulator as a coherent source of hard X rays.
This finding is a crucial milestone in developing a new type of lasers radiating
in the hard X ray and gamma ray range.
\end{abstract}


\maketitle

In this Letter we study for the first time the evolution of a modulated 
particle beam in a planar crystal channel and demonstrate that it preserves 
its modulation at sufficiently large
penetration depths  to ensure the feasibility
of using a crystalline undulator as a coherent source of hard X rays.
Solving this problem is of crucial 
importance in the theory of the Crystal Undulator based Laser (CUL) \cite{first,KGS1999,klystron}
--- a new electromagnetic radiation source in hard X and gamma ray range.

Channeling takes place if charged particles enter a single crystal at small
angle with respect to crystallographic planes or axes \cite{Lindhard}. The 
particles get confined
by the interplanar or axial potential and follow the shape of the corresponding 
planes and axes.

\begin{figure}[ht]
\includegraphics*[width=8.5cm]{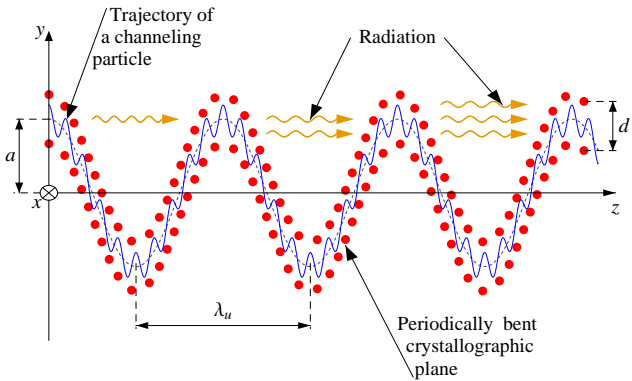}
\caption{Schematic representation of the crystalline undulator.
}
\label{undulator.fig}
\end{figure}

A single crystal with periodically bent crystallographic planes can force 
channeling particles to move along  nearly sinusoidal trajectories
and radiate in hard X and gamma ray frequency range (see Fig. \ref{undulator.fig}).
The feasibility of such a device, known as the 'crystalline undulator`,
was demonstrated theoretically a decade ago \cite{first} (further
developments as well as historical references are reviewed in \cite{KSG2004_review}).
Its experimental study is on the way within the PECU project \cite{PECU}. 

The advantage of the crystalline undulator is in extremely strong
electrostatic fields inside a crystal which are able
to steer the particles much more effectively than even the most advanced
superconductive magnets. 
Due to this fact, such an undulator can radiate powerful electromagnetic waves in
the hard X and soft gamma ray range, where conventional sources with 
comparable intensity are unavailable \cite{Topics}.

Even more powerful and coherent radiation will be emitted if
the probability density of the particles in the beam is modulated
in the longitudinal direction with the period $\lambda$, equal
to the wavelength of the emitted radiation.
In this case, the electromagnetic waves emitted  in the forward direction by
different particles have approximately the same phase \cite{Ginzburg}. Therefore,
the intensity of the radiation becomes proportional
to the beam density squared (in contrast to the linear proportionality for an unmodulated
beam). This increases the photon flux {\it by orders of magnitude}
relative to the radiation of unmodulated beam of the same density.
The radiation of a modulated beam in an undulator is 
a keystone of the physics of free-electron lasers \cite{Madey}.
It can be considered as  a
classical counterpart of the stimulated  emission in quantum physics.
Therefore, if similar phenomenon takes place in a crystalline undulator, it can be
referred to as the {\it lasing regime of the crystalline undulator}.


The feasibility of CUL radiating in hard
X ray and gamma ray range was considered for the fist time in \cite{first,KGS1999}. 
Recently,
a two-crystal scheme, the gamma klystron, has been proposed \cite{klystron}.

\begin{figure}[ht]
\includegraphics[width=8.5cm]{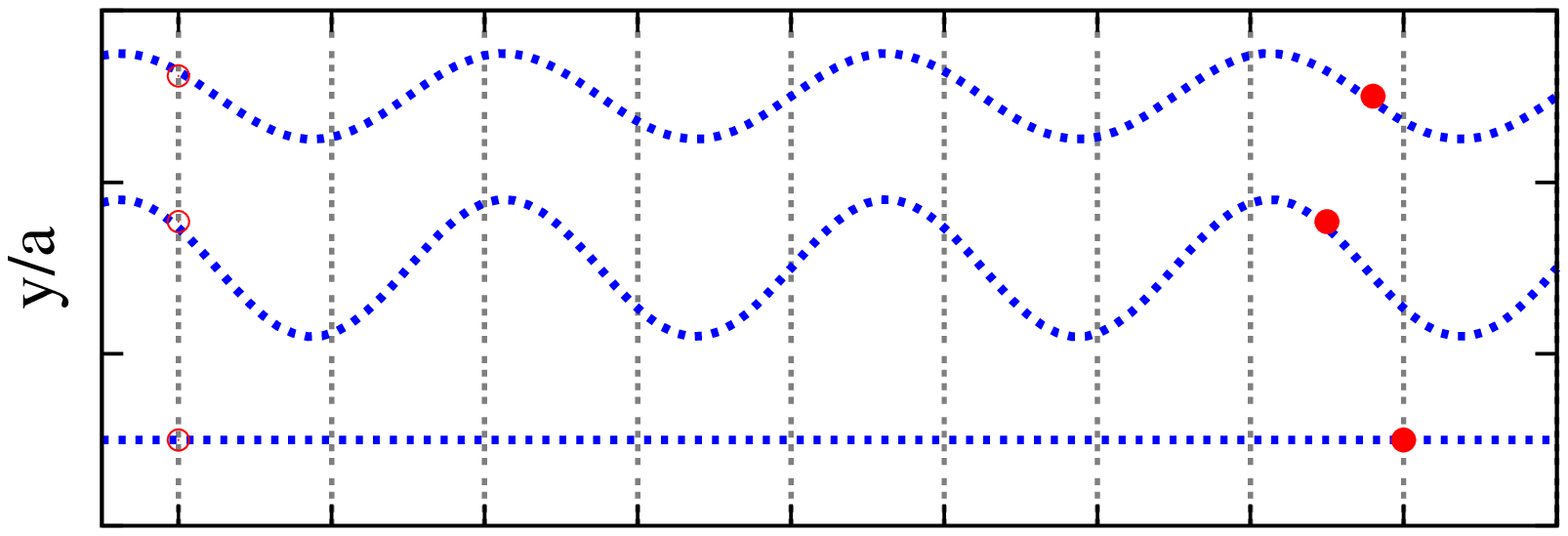}
\includegraphics[width=8.5cm]{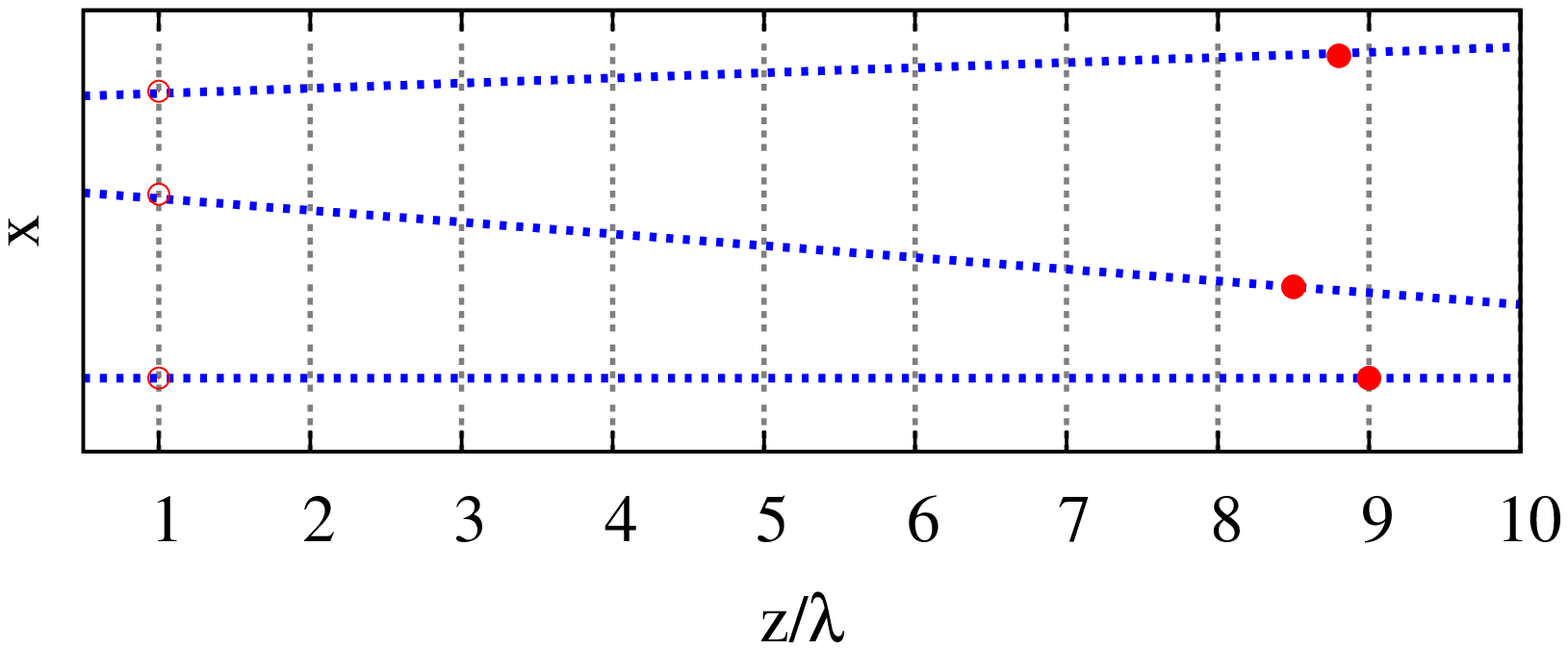}
\caption{Due to different amplitudes of channeling oscillation
(upper panel) and different momentum directions 
in the $(xz)$ plane (lower panel), the initially modulated
beam gets demodulated. Open and filled circles denotes the
same particles at the crystal entrance and after
traveling some distance in the crystal channel, respectively.
}
\label{demodulation.fig}
\end{figure}

A simplified model used in the cited papers assumed that all particle trajectories
follow exactly the shape of the bent channel. In reality, however, the particle
moving along the channel also oscillates in the transverse direction
with respect 
to the channel axis (see the shape of the trajectory 
in Fig. \ref{undulator.fig}). Different particles have
different amplitudes of the oscillations inside the channel
(Fig. \ref{demodulation.fig}, upper panel).
Similarly, the directions of particle momenta
in $(xz)$ plane are slightly different
(Fig. \ref{demodulation.fig}, lower panel).
Even if the speed of the particles along their
trajectories is the same, the particles oscillating with 
different amplitudes or the particles with different trajectory
slopes with respect to $z$ axis have slightly
different $z$ components of their velocities.
As a result, the beam gets demodulated.
An additional contribution to the beam demodulation comes 
from incoherent collisions
of the channeling particles with the crystal constituents.

In the case of an unmodulated beam, 
the length of the crystalline undulator 
and, consequently, the maximum accessible intensity of the radiation
is limited by the dechanneling process.
The channeling particle gradually gains the energy of transverse 
oscillation due to collisions with crystal constituents.
At some point this energy exceeds the maximum value of the interplanar
potential and the particle leaves the channel. The average penetration length 
at which this happens is known as the {\it dechanneling length}.
The dechanneled particle does not follow the sinusoidal 
shape of the channel any more and, therefore, does not contribute to the 
undulator radiation. Hence, 
the reasonable length of the crystalline undulator 
is limited to a few dechanneling lengths. A longer crystal would attenuate rather
then produce the radiation. Since the intensity of the undulator radiation is
proportional to the undulator length squared, the dechanneling length is the 
main restricting factor that has to be taken into account when the radiation output 
is calculated.

In contrast, not only the shape of the trajectory but also the particles positions 
with respect to each other along $z$ axis are important for the lasing regime.
If this positions become random because of the beam demodulation, the intensity
of the radiation drops even if the particles are still in the channeling mode.
Hence, it is the beam demodulation rather than dechanneling that restricts the 
intensity of the radiation of CUL.
Understanding this process and estimating the characteristic length at which this phenomenon
takes place is, therefore, a cornerstone of the theory of this
new radiation source.

Let us consider the distribution 
$f(t,z;\xi,E_{y})$
of the beam particles 
with respect to the angle between the particle trajectory
and axis $z$ in the $(xz)$ plane 
$\xi = \arcsin p_{x}/p \approx p_{x}/p$ and the energy of the channeling
oscillation $E_{y} = p_{y}^2/2 E + U(y)$ \footnote{We chose the system of units in such 
a way that the speed of light is equal to unity.
Therefore, mass, energy and momentum have the same dimensionality. This is also true
for length and time.}. Here $p$, $p_{x}$ and $p_{y}$ are, respectively, the particle momentum and its $x$ and $y$ components,
$U(y)$ is the interplanar potential, and 
$E$ is the particle energy (we will consider only ultrarelativistic 
particles, therefore $E \approx p$).
It can be shown that the
evolution of this distribution in the  crystal
channel with the time $t$ and the
longitudinal coordinate (penetration depth
into the crystal) $z$ can be described by the following differential
equation of Fokker-Planck type:
\begin{equation}
\frac{\partial f}{\partial t} +  
\frac{\partial f}{\partial z} v_{z}
 = D_0 \left [ \frac{\partial }{\partial E_{y}}
\left(E_{y}
\frac{\partial f }{\partial E_{y}}
\right)
+
\frac{1}{E} 
\frac{\partial^2 f }{\partial \xi^2}
\right ] \ .
\label{diffeqD_0}
\end{equation} 
Here $D_0$ is the diffusion coefficient that is dominated 
by the scattering of the beam particles by lattice electrons.
The particle longitudinal velocity
averaged over the undulator period, $v_{z}$,  is given by 
\begin{equation}
v_{z} =    \left( 
1 - \frac{1}{2 \gamma^2} - \frac{\xi^2}{2} - \frac{E_{y}}{2 E}
\right)
\end{equation} 

If the beam is periodically modulated (bunched) the distribution
can be represented as a Fourier series:
\begin{equation}
f(t,z;\xi,E_{y}) = \sum_{j=-\infty}^{\infty} g_{j} (z;\xi,E_{y}) \exp (i j \omega t).
\end{equation}
with $g_{j}^{*} (z;\xi,E_{y}) = g_{-j} (z;\xi,E_{y})$ to ensure the real value of the
particle distribution. Since Eq. (\ref{diffeqD_0}) is linear, 
it is sufficient to consider only one harmonic.
Substituting 
$f(t,z;\xi,E_{y}) = g (z;\xi,E_{y}) \exp (i \omega t)$ one obtains
\begin{equation}
i \omega g(z;\xi,E_{y}) +
 \frac{\partial g}{\partial z} 
v_{z}
 = D_0 \left [ \frac{\partial }{\partial E_{y}}
\left(E_{y}
\frac{\partial g }{\partial E_{y}}
\right)
+
\frac{1}{E} 
\frac{\partial^2 g }{\partial \xi^2}
\right ] .
\label{diffeqg}
\end{equation}

To simplify this equation, we make the substitution
$g(z;\xi,E_{y}) = \exp \left( - i \omega z \right)
\tilde{g}(z;\xi,E_{y})
\label{gtilde}$
and assume that the variation of  
$\tilde{g}(z;\xi,E_{y})$ within the modulation period is small:
$\partial \tilde{g} / \partial z \ll \omega \tilde{g}(z;\xi,E_{y})$.
This allows us to neglect the terms 
$(1-v_{z}) \partial \tilde{g} / \partial z$
while keeping the terms 
$(1-v_{z}) \omega \tilde{g}(z;\xi,E_{y})$.
The resultant partial differential equation
for $\tilde{g}(z;\xi,E_{y})$
can be solved by the method of separation of 
variables. Putting
$\tilde{g}(z;\xi,E_{y}) = \mathcal{Z}(z) \Xi(\xi) \mathcal{E} (E_{y})$,
we obtain a set of ordinary differential equations:
\begin{eqnarray}
\frac{D_0}{E} 
\frac{1}{\Xi(\xi)} 
\frac{d^2 \Xi(\xi)}{d \xi^2} 
- i \omega \frac{\xi^2}{2}
&=& \mathcal{C}_{\xi} , \label{eqXi} \\
\frac{D_0}{\mathcal{E}(E_{y})}
\frac{d }{d E_{y}}
\left(E_{y}
\frac{d  \mathcal{E}(E_{y})}{d E_{y}}
\right) 
- i \omega \frac{E_{y}}{2 E}
&=&  \mathcal{C}_{y}, \label{eqE} \\
\frac{1}{\mathcal{Z}(z)}
\frac{d \mathcal{Z}(z)}{d z} + 
\frac{i \omega}{2 \gamma^2} &=& \mathcal{C}_{z} , \label{eqZ} 
\end{eqnarray}
where $\mathcal{C}_{z}$, $\mathcal{C}_{\xi}$ and $\mathcal{C}_{y}$ do not
depend on any of the variables $z$, $\xi$ and $E_{y}$ and satisfy the condition 
\begin{equation}
\mathcal{C}_{z} = \mathcal{C}_{\xi} + \mathcal{C}_{y} .
\label{sumC}
\end{equation}

Eq. (\ref{eqXi}) has the form of the Schr{\"o}dinger 
equation for the harmonic oscillator.
Its eigenvalues and eigenfunctions are, respectively,
\begin{equation}
\mathcal{C}_{\xi,n}
 = - (1+i)
\sqrt{\frac{\omega D_0}{E}} \left( n + \frac{1}{2} \right),
\ \ \ n = 0,1,2, \dots
\label{Cxi}
\end{equation}
and
\begin{equation}
\Xi_{n}(\xi) =
H_{n} \left( \mathrm{e}^{i \pi / 8} \sqrt[4]{\frac{\omega E}{2 D_0}} \; \xi \right) 
\exp \left( -  \frac{1 + i}{4} \sqrt{\frac{\omega E}{D_0}} 
\xi^2 \right).
\label{Xin}
\end{equation}
Here
$H_{n}(\dots)$ are Hermite polynomials.

Eq. (\ref{eqE}) can be reduced to the Laguerre differential equation, so that 
its solution can be represented as
\begin{equation}
\mathcal{E}_{k}(E_{y})
\! = \!
\exp \!
\left( \!
- \frac{1+i}{2} \sqrt{\frac{\omega}{D_{0} E}} E_{y}
\right)
L_{\nu_{k}} \! \!
\left( \!
(1+i) \sqrt{\frac{\omega}{D_{0} E}} E_{y} \!
\right)
\label{Ek}
\end{equation}
where $L_{\nu}(\dots)$ is the Laguerre function\footnote{At nonnegative integer
values of $\nu$, the Laguerre function is reduced to the well known Laguerre polynomials.
In the general case that is relevant to our consideration, it can be represented by 
an infinite series:
$L_{\nu}(\varepsilon) =  \sum_{j=1}^{\infty} \prod_{m=0}^{j-1} (m - \nu) 
\varepsilon^j / (j!)^2$.} and $\nu_{k}$  is related to the eigenvalue  
$\mathcal{C}_{y,k}$ via
\begin{equation}
\mathcal{C}_{y,k} = 
- \frac{(1+i)}{2} 
\sqrt{\frac{D_{0} \, \omega}{E}} (2 \nu_{k} + 1),
\ \ \ k=1,2,3,\dots
\label{Cynu}
\end{equation}
The eigenvalues can be found by imposing the boundary conditions: the density 
of the channeling particles becomes zero if the energy of channeling oscillations
$E_{y}$ equals to the interplanar barrier $U_{\max}$:
\begin{equation}
L_{\nu_{k}} 
\left( 
(1+i) \sqrt{\frac{\omega}{D_{0} E}} U_{\max}
\right) = 0.
\label{boundcond}
\end{equation}
Equation (\ref{boundcond}) has to be solved  for $\nu_{k}$.
(the subscript $k$ enumerates different roots of the 
equation) and the result has to be substituted into (\ref{Cynu}).

It is convenient to represent the eigenvalues in the form
\begin{equation}
\mathcal{C}_{y,k} = \frac{\alpha_k(\kappa)}{L_\mathrm{d}}  
+ i \omega \theta_\mathrm{L}^2 \beta_k(\kappa).
\end{equation}
Here $L_\mathrm{d} = 4 U_{\max} /(j_{0,1}^2 D_0)$ is the 
dechanneling length \cite{BiryukovChesnokovKotovBook}
($j_{0,k}$  is $k$-th zero of the Bessel function $J_0(\varepsilon)$), 
$\theta_\mathrm{L} = \sqrt{2 U_{\max}/E}$ is Lindhard's angle.
We introduced the parameter
\begin{equation}
\kappa = \pi \frac{L_\mathrm{d}}{\lambda}  \theta_\mathrm{L}^2,
\label{kappa}
\end{equation}
where $\lambda = 2 \pi / \omega$ is the spatial period of the modulation.
The functions $\alpha_k(\kappa)$ and $\beta_k(\kappa)$ are to be found
by solving numerically Eq. (\ref{boundcond}) combined with
(\ref{Cynu}).

Using (\ref{sumC}), one finds the solution 
of Eq. (\ref{eqZ}):
\begin{widetext}
\begin{equation}
\mathcal{Z}_{n,k}(z) = \exp \left \{
- \frac{z}{L_\mathrm{d}} 
\left[ \alpha_k(\kappa) + (2 n + 1) 
\frac{\sqrt{\kappa}}{j_{0,1}}
\right ] 
-
i \omega z
\left [
\frac{1}{2 \gamma^2} + 
\theta_\mathrm{L}^2 \beta_k(\kappa) + 
\theta_\mathrm{L}^2  \frac{(2 n + 1)}{2 j_{0,1} \sqrt{\kappa}}
\right ]
\right \} .
\label{Znk}
\end{equation}
\end{widetext}

Hence, the solution of Eq. (\ref{diffeqg}) is
represented as
\begin{equation}
g(z;\xi,E_{y}) = \exp \left( - i \omega z \right) \sum_{n=0}^{\infty} \sum_{k=1}^{\infty}
\mathfrak{c}_{n,k}
\Xi_{n}(\xi) \mathcal{E}_{k}(E_{y}) \mathcal{Z}_{n,k}(z) ,
\end{equation}
where the coefficients $\mathfrak{c}_{n,k}$ are found from the particle distribution at the
entrance
of the crystal channel.
Due to the exponential decrease of $\mathcal{Z}_{n,k}(z)$ with $z$
(see (\ref{Znk})), the asymptotic behaviour of $\tilde{g}(z;\xi,E_{y})$ 
at large $z$
is dominated by the term with $n=0$ and $k=1$
having the smallest value of the
factor $\left[ \alpha_k(\kappa) + (2 n + 1) \sqrt{\kappa}/j_{0,1}
\right ]$ in the exponential.
Therefore, at sufficiently large penetration depths, the particle
distribution depends on $z$ as
$
g (z;\xi,E_{y}) \propto \exp \left (
- z / L_{\mathrm{dm}}  - i \omega/u_{z} \, z
\right) 
$
where $ L_{\mathrm{dm}}$ is the newly introduced parameter --- {\it the demodulation 
length}:
\begin{equation}
L_{\mathrm{dm}} = \frac{ L_\mathrm{d} }{\alpha_1(\kappa) + \sqrt{\kappa}/j_{0,1}}
\label{Ldm}
\end{equation}
and $u_{z}$ is the phase velocity of the modulated beam along the crystal channel
\begin{equation}
u_{z} = \left [ 1 + \frac{1}{2 \gamma^2} + 
\theta_\mathrm{L}^2 \left( \beta_k(\kappa) + \frac{1}{2 j_{0,1} \sqrt{\kappa}} \right) 
 \right  ]^{-1} .
\label{uz}
\end{equation}
This parameter is important for establishing the resonance conditions between 
the undulator parameters and the radiation wavelength. It
will be analyzed elsewhere.

\begin{figure}[ht]
\includegraphics*[width=8.5cm]{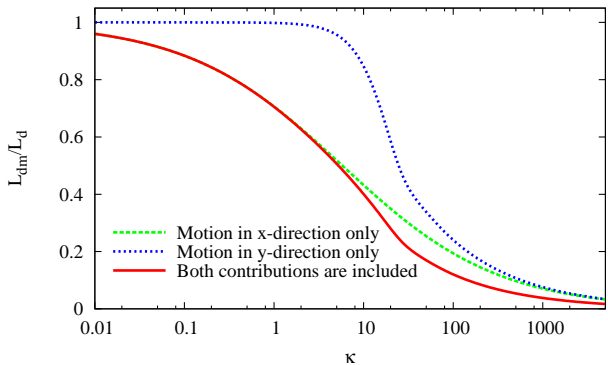}
\caption{The ratio of the demodulation length 
$L_{\mathrm{dm}}$ (\ref{Ldm})
to the dechanneling length $L_\mathrm{d}$ versus 
the parameter $\kappa$ (\ref{kappa}). See text for details.}
\label{Ldm.fig}
\end{figure}

In this Letter we concentrate our attention on the demodulation length.
This parameter represents the characteristic scale of the penetration depth at
which a beam of channeling particles gets demodulated.
Fig. \ref{Ldm.fig} presents the dependence of the ratio $L_{\mathrm{dm}}/L_\mathrm{d}$ on
the parameter $\kappa$. It is seen that the demodulation 
length approaches the
dechanneling length at $\kappa \lesssim 1$. This is an important result, since
it means that the demodulation process
does not put additional restrictions on the undulator length and, therefore, 
a strong lasing effect can be expected in the crystalline 
undulator fed by a modulated beam.
On the contrary, the ratio noticeably drops for $\kappa \gtrsim 10$.
Therefore, further investigations are needed to clarify whether the lasing regime
is feasible for these $\kappa$.

\begin{figure}[ht]
\includegraphics*[width=8.5cm]{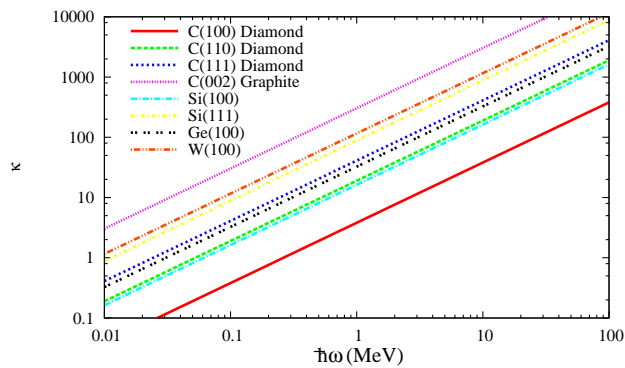}
\caption{The parameter $\kappa$ (\ref{kappa}) versus the photon energy
$\hbar \omega$ for different crystals and crystallographic 
planes.}
\label{hbaromega_kappa.fig}
\end{figure}

The dependence of the parameter $\kappa$ on the energy of the emitted photons, 
$\hbar \omega = 2 \pi \hbar/\lambda$, is shown in Fig. \ref{hbaromega_kappa.fig}.
The calculation was done for 1 GeV\footnote{Note that $\kappa$
depends weakly (logarithmically) on the particle energy. Therefore,
changing the beam energy by an order of magnitude would leave
Fig. \ref{hbaromega_kappa.fig} practically unchanged.} positrons using the formula 
for the dechanneling length from \cite{KSG2004_review,BiryukovChesnokovKotovBook}.

As one sees from the figure, there exist such crystal channels  that $\kappa = 1$
corresponds to $\hbar \omega = 100-300$ keV. Therefore, CUL is most 
suitable for the application in this photon energy range.
Using crystalline undulator for the emission of softer photons $\hbar \omega \lesssim 10$ keV
is prevented by very strong absorption of the radiation in the crystal, while the upper
limit on the photon energy is set by the decreasing value of $L_{\mathrm{dm}}$.

It is instructive to study the influence of the particle motion in $x$ and $y$ direction
on the demodulation length separately. Replacing $\alpha_1(\kappa)$ in (\ref{Ldm}) with unity  
means neglecting the motion in the $y$ direction, while omitting the second term in the denominator
ignores the motion in $x$ direction. One sees from Fig. \ref{Ldm.fig} that
it is mostly the motion in $x$ direction that suppresses the demodulation length
at $\kappa \lesssim 10$,
while the influence of channeling oscillations is negligible. This suggests the idea
that axial channeling, i.e. when motion in both $x$ and $y$ directions has the nature 
of channeling oscillations, might be more suitable for the lasing regime
in the range 
$\hbar \omega \sim 1$ MeV.

In conclusion, we have studied the propagation of a modulated particle beam in a planar
crystal channel.
It has been demonstrated that the beam preserves its modulation at the penetration depths
which are sufficient for using the crytalline undulator as a source of coherent radiation 
with the photon energy of hundreds keV.

We restricted our analysis to the beam demodulation in a straight crystal. The influence 
of the periodic bending on the demodulation length has to be studied at the next step.
Another important milestone in the theory of CUL would be developing suitable
methods of beam modulation.

This work has been supported by the European Commission 
(the PECU project, Contract No. 4916 (NEST)).



\begin{thebibliography}{99}
\bibitem{first}
         A.V.~Korol, A.V.~Solov'yov, W.~Greiner, 
         J. Phys. G {\bf 24}, L45 (1998); 
\bibitem{KGS1999}
         Int. J. Mod. Phys. E {\bf 8} 49 (1999).
\bibitem{klystron}   
A.~Kostyuk, A.~Korol, A.~Solov'yov and W.~Greiner,
arXiv:0710.4772 [physics.acc-ph].
\bibitem{Lindhard}
         J.~Lindhard, Kong. Danske Vid. Selsk. Mat.-Fys. Medd.
        {\bf 34}, 14 (1965).
\bibitem{KSG2004_review}
         A.V.~Korol, A.V.~Solov'yov,  W.~Greiner, 
         Int. J. Mod. Phys. E {\bf 13}, 867 (2004).
\bibitem{PECU}
http://ec.europa.eu/research/fp6/nest/pdf

\bibitem{Topics}
A.V.~Korol, A.V.~Solov'yov,  W.~Greiner, 
Topics in Heavy Ion Phys., 73 (2005).

\bibitem{Ginzburg}
V.L.~Ginzburg, Izv. Akad. Nauk. SSSR, Ser. Fiz. {\bf 11}, 165 (1947).

\bibitem{Madey}
J.M.J.~Madey, J. Appl. Phys. {\bf 42}, 1906 (1971).

\bibitem{BiryukovChesnokovKotovBook}
         V.M.~Biruykov,  Yu.A.~Chesnokov, V.I.~Kotov,
        {\it Crystal Channelling and its Application at 
         High-Energy  Accelerators}
         (Springer, Berlin, 1996).

\end{thebibliography}
\end{document}